\documentclass[pra, preprint,  showpacs,amssymb]{revtex4}
\usepackage{amsmath}
\usepackage{subfigure}
\usepackage{latexsym}
\usepackage{amsmath}
\usepackage{amsfonts}
\usepackage{graphicx}
\usepackage{fancyhdr}
\usepackage{epsfig}
\usepackage{graphics}
\begin{document}
\title{Motion of position-dependent mass  as a damping-antidamping process: Application to the Fermi gas and to the  Morse potential}

\author{Yamen Hamdouni}\email[Email:\ ]{ hamdouniyamen@gmail.com}\address{ 104 Deuxi\`{e}me  Tranche, Bekira, Constantine, Algeria} 

\begin{abstract}
The object of this paper is to investigate, classically and quantum mechanically, the relation existing between the position-dependent effective mass and  damping-antidamping dynamics.  The quantization of the equations of motion is carried out  using the geometric interpretation of the motion, and we compare it with the one based on the ordering ambiguity scheme. Furthermore,  we apply the obtained results to a Fermi gas of damped-antidamped particles, and we solve the Schr\"odinger equation for an exponentially increasing (decreasing) mass in the presence of the Morse potential.
\end{abstract}


\maketitle

\section{Introduction}
  In the course of the development of the quantum theory, Schr\"odinger's equation has played a central role~\cite{land}. It represents the basic tool for studying the behavior of  many  quantum systems ranging from nuclei~\cite{greiner} to solid-state compounds~\cite{kittel}.  Very often, one deals with systems whose masses are constant in time and in space. Fortunately, this leads in many cases to exact results regarding the energy spectra and the eigenfunctions of the associated Hamiltonians, as is the case for, among others, the harmonic oscillator and the hydrogen atom.   
	
	There exist, on the other hand, physical systems for which the mass exhibits the property of being time and/or position dependent. In particular, the motion of position-dependent effective mass (PDEM) has been the subject of a great deal of investigations, owing to its  connection with the description of various physical systems such as quantum dots~\cite{dots,dots2}, semiconductors~\cite{semi,semi1,semi2,semi3,semi4,semi5,semi6,semi7}, metal clusters~\cite{metal}, ${\rm ^{3}He}$ clusters~\cite{he}, and  quantum liquids~\cite{liquid}. Depending on the intended application,	many techniques have been used in order to study the PDEM motion~\cite{chro,chro1,chro2,chro3,chro4,chro5,chro6,chro7,chro8,chro9, chro10,chro11,chro12, chro13,chro14,chro15}; the standard approach consists in solving the  Schr\"odinger equation,  where one inevitably has to deal with the nontrivial ordering problem of the momentum operator and the mass in the kinetic energy part of the PDEM Hamiltonian~\cite{ord}.  
		
	In this paper, we explore  the  PDEM motion from a dynamical point of view, and we investigate its  damping-antidamping character  both classically and quantum mechanically. Generally speaking, the damping of classical particles is  described by a frictional force that is  proportional to the velocity (i.e., a force which is an odd function of the velocity)~\cite{odd}. In effect,  the system under investigation is dissipative and is, quite often, non-Hamiltonian, making it difficult to quantize the equations of motion. This is the reason for which different approaches are adopted in the attempt to achieve  a satisfactory quantum mechanical description of damped systems~\cite{weiss,gardiner}.  Moreover, frictional forces that depend on the square of the velocity are known from hydrodynamics. As we shall see latter, the PDEM motion is tightly related to  such forces, which may display antidamping behavior due to the fact that they are even functions of the velocity. The question that naturally arises is whether the corresponding equations of motion can be quantized using the standard technique. We shall be concerned with the answer of this question in the  first part of this paper.
	
	The Fermi gas has found many applications in various fields of physics, like nuclear physics~\cite{hofman}, and solid-state physics~\cite{kittel}. 
Hence it is of interest to investigate the consequences of the spatial dependence of the mass  on the properties of this system. 
 In this spirit, we feel tempted to  extend the analysis to the molecular domain, by studying the PDEM motion under the effect of the Morse potential~\cite{morse}. 
	
	The paper is organized as follows. In Section~{\ref{sec2}} we establish the relation between the PDEM  and damping-antidamping motion. There, the quantization of the equations of motion is carried out through the  geometric interpretation of the dynamics, which enables us to find the stationary states corresponding to quasi-free particles for particular forms of the  spatial dependence of the mass.  Section~\ref{sec3} is devoted to the application of the obtained results to a gas of damped-antidamped particles at zero and nonzero temperatures; the emphasis will be on the consequences on the pressure and the specific heat of the gas. In Section~\ref{sec4}, we solve the Schr\"odinger equation when the particle is endowed with exponentially decreasing (or increasing) mass, under the effect of  the Morse potential, to determine the energy spectra and the eigenfunctions. The paper is ended with a summary.

\section{Damping-antidamping interpretation of the PDEM motion\label{sec2}}
\subsection{Classical treatment}
In one dimension, the Hamiltonian  corresponding to a position-dependent effective mass $m(q)=m_0 M(q)$, where $M(q)$ is  a dimensionless positive function of the generalized coordinate $q$, is given by
\begin{equation}
H(p,q)=\frac{p^2}{2 m(q)}+V(q)=\frac{p^2}{2 m_0 M(q)}+V(q).\label{ham1}
\end{equation}
Here $V(q)$ is the potential, and $p$ denotes the generalized canonical  momentum conjugate to $q$, that is:
\begin{equation}
\{q,p\}=1,\qquad \{q,q\}=\{p,p\}=0,
\end{equation}
where the curly braces denote Poisson Brackets. Using the canonical equations of  motion
\begin{eqnarray}
\dot{q}&=&\frac{\partial H}{\partial p},\nonumber\\
\dot{p}&=&-\frac{\partial H}{\partial q},
\end{eqnarray}
it can  easily be shown that
\begin{equation}
\dot{p}=\frac{1}{2} [\nabla_q m(q)] {\dot q}^2-\nabla_q V(q),
\end{equation}
where $\nabla_q\equiv\partial_q$, and the dot designates the derivation with respect to time.  Taking into account the  fact that $\dot m(q)=\dot q \nabla_qm(q)$,  it immediately follows that
\begin{equation}
\ddot{q} =-\frac{1}{2 m(q)}[\nabla_q m(q)]{\dot q}^2-\frac{1}{m(q)}\nabla_q V(q).
\end{equation}
Now multiplying both sides of the latter equation by the constant  $m_0$ yields
\begin{equation}
m_0\ddot{q} =-\Bigl[\frac{m_0}{2 M(q)}\nabla_q M(q)\Bigr]{\dot q}^2-\nabla_q\Biggl\{\int\limits_q\frac{1}{M(q)}\nabla_q V(q)dq+V_0\Biggr\},\label{motion}
\end{equation}
where $V_0$ is a constant. Clearly,~(\ref{motion}) is the equation of motion corresponding to a particle of constant mass, $m_0$, moving in  one dimension, subjected to the effect of two forces; the first one being proportional to the square of the velocity:
\begin {equation}
F_{d}=-\Bigl[\frac{m_0}{2 M(q)}\nabla_q M(q)\Bigr]{\dot q}^2,\label{fric}
\end{equation} while the second one is a conservative force, derived from the effective potential
\begin{equation}
V_{\rm eff}(q)= \int\limits_q\frac{1}{M(q)}\nabla_q V(q)dq+V_0.\label{fric4}
\end{equation}
 Integrating by parts, we can rewrite $V_{\rm eff}$ as
\begin{equation}
V_{\rm eff}(q)=\frac{V(q)}{M(q)}+\int\limits_q \frac{V(q)}{M(q)^2}\nabla_q M(q)dq+V_0.
\end{equation}

Let us now take a look at equation~(\ref{fric}). It is clear that $F_d$ is an even function of the generalized velocity $\dot{q}$, meaning that the nature of the force is dictated by the explicit forms of  $M(q)$ and $V(q)$. Suppose that the latter is such that  the particle is constrained to move in some definite region of  space. Consequently, $F_d$ is of frictional nature (the particle is decelerated) when  $\dot q>0$ and $\nabla_q M(q)>0$ or $\dot q<0$ and $\nabla_q M(q)<0$. On the contrary the particle gains acceleration (antidamping) if $\dot q>0$ and $\nabla_q M(q)<0$ or $\dot q<0$ and $\nabla_q M(q)>0$. This may be better seen by noting that $\nabla_q M(q)=\dot M(q)/\dot q$. When $V(q)$ is zero and $M(q)$ monotonic,  the particle (which we call quasi-free) is either damped or antidamped depending on its initial velocity. We thus come to the conclusion that the PDEM motion is equivalent to the damping-antidamping dynamics of a particle with constant mass subjected to a force proportional to the square of the velocity.

In what follows we shall prove the converse. To this end, consider  the following equation of motion
\begin{equation}
m_0 \ddot q=-\phi(q) \dot{q}^2-\nabla_qU(q;\alpha),\label{fric2}
\end{equation}
where $U(q;\alpha)$ is a potential depending on some parameter $\alpha$. Our task here is to find the Hamiltonian  from which equation~({\ref{fric2}) is derived.  First of all
notice that a first integral for~(\ref{fric2}), which we denote by $I$, is a function of $q$ and $\dot q$,  satisfying
\begin{equation}
\frac{d I(q,\dot q)}{dt}=\dot q\frac{\partial I}{\partial q}+\ddot q\frac{\partial I}{\partial \dot q}=0.
\end{equation}
This means that
\begin{equation}
\dot q\frac{\partial I}{\partial q}-\frac{1}{m_0}\Bigl[\phi(q) \dot{q}^2+\nabla_qU(q;\alpha)\Bigr]\frac{\partial I}{\partial \dot q}=0.\label{par1}
\end{equation}
Let $\mathcal C$ denote the family of curves in the $(q,\dot q)$ plane, for which $dI/dq=0$ holds. Hence along these curves we have that
\begin{equation}
\frac{\partial I}{\partial q}+\frac{d\dot q}{ dq} \frac{\partial I}{\partial \dot q}=0.\label{par2}
\end{equation}
 Comparing~(\ref{par1}) and~(\ref{par2}), we get the following first-order ordinary differential equation
 \begin{equation}
 \frac{d\dot q}{dq}=-\frac{1}{m_0}\phi(q)\dot q-\frac{1}{m_0\dot q}\nabla_q U(q;\alpha).\label{par3}
 \end{equation}
 We seek a solution in the form
 \begin{equation}
 \dot q=f(q){\rm e}^{-\frac{1}{m_0}\int_q \phi(q) dq}
 \end{equation}
 where $f(q)$ is some function of $q$. By inserting the proposed solution into~(\ref{par3}), we get after some algebra
 \begin{equation}
 \frac{df^2}{dq}=-\frac{2}{m_0}{\rm e}^{\frac{2}{m_0}\int_q \phi(q) dq}\nabla_q U(q;\alpha).
 \end{equation}
This can easily be integrated, and we get
 \begin{equation}
 f(q)^2=-\frac{2}{m_0}\int\limits_q{\rm e}^{\frac{2}{m_0}\int_q \phi(q) dq}\nabla_q U(q;\alpha) dq+C.
 \end{equation}
Here $C$ is a constant of integration characteristic of the curves $\mathcal C$. Finally, the general solution of~(\ref{par3}) is
 \begin{equation}
 \dot q^2=C {\rm e}^{-\frac{2}{m_0}\int_q \phi(q) dq}-\frac{2}{m_0}{\rm e}^{-\frac{2}{m_0}\int_q \phi(q) dq}\int\limits_q{\rm e}^{\frac{2}{m_0}\int_q \phi(q) dq}\nabla_q U(q;\alpha) dq.
 \end{equation}
 We solve for $C$ to obtain
 \begin{equation}
 C(q,\dot q)=\dot q^2{\rm e}^{\frac{2}{m_0}\int_q \phi(q) dq}+\frac{2}{m_0}\int\limits_q{\rm e}^{\frac{2}{m_0}\int_q \phi(q) dq}\nabla_q U(q;\alpha) dq.
 \end{equation}
 The simplest choice we can make for the first integral of the equation of motion is
 \begin{equation}
 I(q,\dot q)=\frac{1}{2}m_0 g(\alpha) C(q,\dot q)+K(\alpha),
 \end{equation}
 where $g(\alpha)$ is a dimensionless positive function of the parameter $\alpha$ and where $K(\alpha)$ has the dimension of energy.

According to~\cite{int1,int2,int3,int4},  the Lagrangian of the system is  given in terms of the first integral by
 \begin{equation}
 L=\dot q\int\frac{I(q,\dot q)}{\dot q^2} d\dot q.
\end{equation}
A straightforward calculation yields
\begin{equation}
 L=\frac{1}{2}m_0g(\alpha)\dot q^2 {\rm e}^{\frac{2}{m_0}\int_q \phi(q) dq}-g(\alpha)\int\limits_q{\rm e}^{\frac{2}{m_0}\int_q \phi(q) dq}\nabla_q U(q;\alpha) dq-K(\alpha).
 \end{equation}
 It is now easy to show that the generalized momentum is equal to
 \begin{equation}
 p=\frac{\partial L}{\partial \dot q} =m_0 g(\alpha){\rm e}^{\frac{2}{m_0}\int_q \phi(q) dq}\dot q.
 \end{equation}
 Consequently, the Hamiltonian  corresponding to~(\ref{fric2}) is given by
 \begin{eqnarray}
 H(q,p)&=&\dot q p-L\nonumber\\
 &=&\frac{p^2}{2m_0 g(\alpha)}{\rm e}^{-\frac{2}{m_0}\int_q \phi(q) dq}+g(\alpha)\int\limits_q{\rm e}^{\frac{2}{m_0}\int_q \phi(q) dq}\nabla_q U(q;\alpha) dq+K(\alpha).
 \end{eqnarray}
 Clearly, this is the Hamiltonian  for a position-dependent mass
 \begin{equation}
 m(q)=m_0 g(\alpha){\rm e}^{\frac{2}{m_0}\int_q \phi(q) dq}
 \end{equation}
 subjected to the effective potential
 \begin{equation}
 U_{\rm eff}(q;\alpha)=g(\alpha)\int\limits_q{\rm e}^{\frac{2}{m_0}\int_q \phi(q) dq}\nabla_q U(q;\alpha) dq+K(\alpha).\label{eff2}
 \end{equation}

 As an example consider  the motion of the particle under the effect of the force $-\eta \dot q^2$ ($\phi(q)=\eta>0$), in the presence of the Morse potential~\cite{morse}
 \begin{equation}
 U(q;\alpha)=A({\rm e}^{-2\alpha q}-2{\rm e}^{-\alpha
 q})\label{morsa}
 \end{equation}
 with $A,\alpha >0$. From equation~(\ref{eff2}), it follows that when $\alpha\neq\eta/m_0$ and
$\alpha\neq 2\eta/m_0$,
 \begin{equation}
 U_{\rm eff}(q;\alpha)=
   g(\alpha)A\Biggl[\frac{\exp\{2(\tfrac{\eta}{m_0}-\alpha)q\}}{1-\tfrac{\eta}{m_0\alpha}}-\frac{2\exp\{(\tfrac{2\eta}{m_0}-\alpha)q\}}{1-\tfrac{2\eta}{m_0\alpha}}\Biggl] +K(\alpha).\label{morse1}
\end{equation}
On the other hand if $\alpha=\eta/m_0$, then
\begin{equation}
U_{\rm eff}(q;\alpha)=
   2 g(\alpha)A \Bigl[\exp(2\alpha q )-\alpha q \Bigr]+K(\alpha),
   \end{equation}
whereas for $\alpha=2\eta/m_0$ we find that
\begin{equation}
U_{\rm eff}(q;\alpha)=
   2 g(\alpha)A \Bigl[\exp(-2\alpha q )+\alpha q \Bigr]+K(\alpha).
   \end{equation}
   Obviously, $g(\alpha)$ and $K(\alpha)$ are not the same for all the cited circumstances; note, however, that while $K(\alpha)$ may arbitrarily be chosen (additive constant), $g(\alpha)$ should be reduced to unity when $\eta\to0$. Even though, the choice one can make for  $ g(\alpha)$  still seems to be  not unique. It should be stressed that, generally speaking, different Hamiltonians describing the same classical system may yield completely different quantum behaviors. 
	
To illustrate the importance of the inclusion of the function $g$ we calculate  the minimum of the effective potential $U_{\rm eff}$ appearing in equation~(\ref{morse1}):  \begin{equation}
 U_{\rm eff}^{\rm min}=A g(\alpha) \Biggl[\frac{1}{1-\tfrac{\eta}{m_0\alpha}}-\frac{2}{1-\tfrac{2\eta}{m_0\alpha}}\Biggl] +K(\alpha). \label{min}
 \end{equation}
 We  observe that the quantity $Q=\frac{1}{1-\tfrac{\eta}{m_0\alpha}}-\frac{2}{1-2\tfrac{\eta}{m_0\alpha}}$  blows up in the neighborhoods of  $\alpha=\eta/m_0$ and $\alpha= 2\eta/m_0$. Indeed, $Q\to-\infty$ when $\eta/(m_0\alpha)\to 1^+$ or $2\eta/(m_0\alpha)\to 1^-$, whereas $Q\to+\infty$ when $\eta/(m_0\alpha)\to 1^-$ or $2\eta/(m_0\alpha)\to 1^+$, which, obviously, is physically unacceptable.  The only way to remove these singularities  consists in choosing the following form for the function $g(\alpha)$:
 \begin{equation}
 g(\alpha)=\widetilde g(\alpha)\Bigl|(1-\tfrac{\eta}{m_0\alpha})(1-\tfrac{2\eta}{m_0\alpha})\Bigr|
 \end{equation}
 where $\widetilde g(\alpha)$ is a regular positive dimensionless function of $\alpha$. Therefore the effective potential reads
 \begin{eqnarray}
 U_{\rm eff}(q;\alpha)&=&   \widetilde g(\alpha)A\Biggl[{\rm{sgn}}\Bigl(1-\tfrac{\eta}{m_0\alpha}\Bigr)\Bigl|1-\tfrac{2\eta}{m_0\alpha}\Bigr|\exp\Bigl\{2(\tfrac{\eta}{m_0}-\alpha)q\Bigr\}\nonumber\\ &-&2\ {\rm{sgn}}\Bigr(1-\tfrac{2\eta}{m_0\alpha}\Bigr)\Bigl|1-\tfrac{\eta}{m_0\alpha}\Bigr|\exp\Bigl\{(\tfrac{2\eta}{m_0}-\alpha)q\Bigr\}\Bigr]+K(\alpha).
\end{eqnarray}
As a result,  the mass becomes $\alpha$ dependent.

\subsection{Quantization}
From a geometric point of view, the Hamiltonian~(\ref{ham1}) describes, strictly speaking, the motion of a particle with constant mass $m_0$ in a one-dimensional space whose metric is defined by
\begin{equation}
ds^2=M(q) dq^2.
\end{equation}
The corresponding Hamiltonian operator reads
\begin{equation}
\hat H=-\frac{\hbar^2}{2m_0}\Delta+ V(q),\label{ham2}
\end{equation}
where $\Delta$ denotes the Laplacian operator. From the differential geometry we know that the Laplacian operator in  the curvilinear coordinates $x_\mu$ with metric  tensor $g_{\mu\nu}$ is given by
\begin{equation}
 \Delta=\frac{1}{\sqrt{{\rm det (g_{\mu\nu})}}}\sum\limits_{\mu,\nu}\frac{\partial}{\partial x_\mu} \sqrt{{\rm det (g_{\mu\nu})}}g_{\mu\nu}^{-1}\frac{\partial}{\partial x_\nu}.
 \end{equation}
 It follows that the time-dependent Schr\"odinger equation, in our case, takes the form
 \begin{equation}
 \Biggl[-\frac{\hbar^2}{2m_0\sqrt{M(q)}}\frac{\partial}{\partial q}\frac{1}{\sqrt{M(q)}}\frac{\partial}{\partial q}+ V(q)\Biggr]\psi(q,t)=i\hbar\frac{\partial\psi(q,t)}{\partial t}.\label{chro}
 \end{equation}
In turn the wave function $\psi(q,t)$ should be normalized according to
\begin{equation}
\int_{-\infty}^{\infty}|\psi(q,t)|^2\sqrt{M(q)}dq=1,{\label{norm}}
\end{equation}
 which actually arises from the fact that the line element  is equal to
 $\sqrt{ M(q)}dq$.
 On the other hand we can write:
 \begin{eqnarray}
 i\hbar\Bigl(\psi^*\frac{\partial\psi}{\partial t}+\psi\frac{\partial\psi^*}{\partial t}\Bigr)&=& -\frac{\hbar^2}{2\sqrt{m(q)}}\Biggl[\frac{\psi^*}{\sqrt{m(q)}}\Delta_q\psi+\frac{\nabla_q\psi\nabla_q\psi^*}{m(q)^{1/2}}-\frac{\nabla_q\psi\nabla_q\psi^*}{m(q)^{1/2}}\nonumber\\&-&\frac{1}{2}\frac{\nabla_q m(q)}{m(q)^{3/2}}\psi^*\nabla_q\psi-\frac{\psi}{\sqrt{m(q)}}\Delta_q\psi^*+\frac{1}{2}\frac{\nabla_q m(q)}{m(q)^{3/2}}\psi\nabla_q\psi^*\Biggr]\nonumber\\ &=&-\frac{\hbar^2}{2\sqrt{m(q)}}\nabla_q\Biggl(\frac{\psi^*}{\sqrt{m(q)}}\nabla_q\psi-\frac{\psi}{\sqrt{m(q)}}\nabla_q\psi^*\Biggr),
 \end{eqnarray}
 where $\nabla_q\equiv\partial_q$, and $\Delta_q\equiv\partial^2_q$. Now by noting that the gradient in the $q$ direction is ${\rm grad}_q\equiv \widetilde{\nabla}_q\equiv M(q)^{-1/2}\nabla_q$, the equation of local conservation of probability  takes the form
 \begin{equation}
 \frac{\partial|\psi|^2}{\partial t}+\widetilde{\nabla}_q  j=0,
 \end{equation}
 with the probability current 
 \begin{equation}
  j=\frac{\hbar}{2im_0}[\psi^*(q,t)\widetilde{\nabla}_q\psi(q,t)-\psi(q,t)\widetilde{\nabla}_q\psi^*(q,t)].
 \end{equation}
 Moreover, if we assume that $m(q)$ does not depend explicitly on time, then  we get another  continuity equation in the usual Cartesian space:
  \begin{equation}
 \frac{\partial\tilde\rho}{\partial t}+\nabla_q \tilde{  j}=0,
 \end{equation} where
 \begin{equation}
 \tilde\rho(q,t)=|\psi(q,t)|^2\sqrt{M(q)}=|\psi(q,t)M(q)^{1/4}|^2,
 \end{equation}
 and where
  \begin{eqnarray} \tilde{ j}(q,t)&=&\frac{\hbar}{2im_0\sqrt{M(q)}}\Biggl[\psi^*\nabla_q\psi-\psi\nabla_q\psi^*\Biggr]\nonumber\\&=&\frac{\hbar}{2im_0}\Biggl[\frac{\psi^*}{M(q)^{1/4}}\nabla_q\Biggl(\frac{\psi}{M(q)^{1/4}}\Biggr)-\frac{\psi}{M(q)^{1/4}}\nabla_q\Biggl(\frac{\psi^*}{M(q)^{1/4}}\Biggr)\Biggr].\label{curr}
	\end{eqnarray}
Hence it is constructive to define a new wave function
\begin{equation}
\varphi=\frac{\psi}{M(q)^{1/4}}.
\end{equation}
After some algebra, it can be shown that the latter wave function satisfies the Schr\"odinger equation
\begin{equation}
-\frac{\hbar^2}{2m_0M(q)}\frac{\partial^2\varphi}{\partial q^2}+{\widetilde V}(q)\varphi=i\hbar\frac{\partial \varphi}{\partial t},
\end{equation} 
where the effective potential is
\begin{equation}
\widetilde V (q)=V(q)+\frac{\hbar^2}{8m_0M(q)}\Biggl[\frac{5(\nabla_q M(q))^2}{4M(q)^2}-\frac{\Delta_q M(q)}{M(q)}\Biggr].
\end{equation}

 \subsubsection*{\bf Case of a quasi-free particle in one dimension}
 Suppose that the potential $ V(q)$ is identically equal to zero in the whole space, and that the position dependence of the mass of the particle is given by
 \begin{equation}
 m(q)=m_0\ {\rm e}^{ -\frac{2\eta}{m_0}q}.\label{mass}
 \end{equation}
 We shall   look  for the stationary states satisfying  the time-independent  Schr\"odinger equation
 \begin{equation}
\hat H\psi(q)=E\psi(q)\label{miss},
 \end{equation}
 where $E>0$ is the energy of the particle. Taking into account~(\ref{chro}) we get
   \begin{equation}
 \frac{d^2\psi}{dq^2}+\frac{\eta}{m_0}\frac{d\psi}{dq}+\frac{2m_0E}{\hbar^2}{\rm e}^{-(2\eta/m_0)q}\psi=0,\label{quasifree1}
 \end{equation}
   By the  change of variable $\xi={\rm e}^{-(\eta/m_0)q}$, the above equation reduces to the  simple second-order differential equation
 \begin{equation}
  \frac{d^2\psi}{d\xi^2}+\frac{2 m_0^3 E}{\hbar^2\eta^2}\psi=0,
  \end{equation}
  which can easily be solved. Hence the general solution of~(\ref{quasifree1}) is
  \begin{eqnarray}
  \psi(q)&=&C_1(\eta)\exp\Biggl\{i\sqrt{\frac{2 m_0^3 E}{\hbar^2\eta^2}} {\rm e}^{-(\eta/m_0)q}\Biggr\}+C_2(\eta)\exp\Biggl\{-i\sqrt{\frac{2 m_0^3 E}{\hbar^2\eta^2}} {\rm e}^{-(\eta/m_0)q}\Biggr\}\nonumber\\
  &=&C_1(\eta)\psi^-(q)+C_2(\eta)\psi^+(q).\label{fun1}
  \end{eqnarray}
The meaning of the notations $\psi^\pm(q)$ is clear from the above equation. We redefine the constants $C_1$ and $C_2$ in such a way that $\psi$ takes the form (assume $\eta>0$)
  \begin{eqnarray}
  \psi(q)&=&C_1(\eta)\exp\Biggl\{-i\sqrt{\frac{2 m_0^3 E}{\hbar^2\eta^2}} (1-{\rm e}^{-\frac{\eta}{m}q})\Biggr\}+C_2(\eta)\exp\Biggl\{i\sqrt{\frac{2 m_0^3 E}{\hbar^2\eta^2}} (1-{\rm e}^{-\frac{\eta}{m}q})\Biggr\}.
  \end{eqnarray}
  Now  the limit of constant mass ($\eta\to0$) may be obtained by a Taylor expansion of the exponential function; this yields
  \begin{equation}
  \lim\limits_{\eta\to0}\psi(q)=\widetilde C_1\ {\rm e}^{-ik q }+ \widetilde C_2\  {\rm e}^{ik q },
  \end{equation}
  where $k=\sqrt{2m_0E/\hbar^2}$ and where $\widetilde C_{1,2}=\lim_{\eta\to0}C_{1,2}(\eta)$. It is  clear that the above equation represents the  usual expression of the wave function for a free particle.

   Consequently, we can interpret $\psi^+(q)$ as the wave function of the particle when it is moving  to the right, whereas $\psi^-(q)$ describes the motion in the opposite direction. Indeed,   it can easily be verified that the probability current density corresponding to $\psi^\pm$ is equal to $\pm\sqrt{2m_0 E}/(\hbar)$. Since $\nabla_q m(q)<0$, we conclude that $\psi^-$ corresponds to the damping of the particle; on the contrary, $\psi^+$ describes the antidamping behavior of our quantum system. 

 Now let us examine the transmission and the reflection of the particle from the rectangular potential wall $U(q)=U_0 \theta(q-a)$ where $U_0>0$ and $\theta(q)$ is the heaviside function. We further assume, for the sake of generality, that the mass of the particle is given by
 \begin{equation}m(q)=m_0\exp\{-2(\eta_1\theta(a-q)+\eta_2\theta(q-a)) q/m_0\}.
 \end{equation}
   Note that the solution of the Schr\"odinger equation when the potential is  $U(q)=U_0={\rm const}$ can be obtained by simply replacing $E$ in~(\ref{fun1}) by $E-U_0$.

First we need to determine the conditions that should be satisfied by the  wave function and its derivative   at the discontinuity point $a$. To this end let us multiply both sides of equation~(\ref{miss}) by $\sqrt{M(q)}$ and integrate between $a-\epsilon$ and $a+\epsilon$, where $\epsilon$ is a small number; we find
\begin{equation}
-\frac{\hbar^2}{2m_0}\lim\limits_{\epsilon\to 0}\Biggl[\frac{1}{\sqrt{M(q)}}\frac{d\psi}{dq}\biggl|_{a+\epsilon}-\frac{1}{\sqrt{M(q)}}\frac{d\psi}{dq}\biggl|_{a-\epsilon}\Biggr]=\lim\limits_{\epsilon\to 0}\int_{a-\epsilon}^{a+\epsilon}\bigl[E-U(q)\bigr]\sqrt{M(q)}dq.
\end{equation}
The right-hand side of the latter equation is identically equal to zero since the quantity under the integral sign is bounded (we assume that $M(q)$ is bounded in the interval $[a-\epsilon, a+\epsilon])$. Hence
\begin{equation}
\Biggl(\frac{1}{\sqrt{M(q)}}\frac{d\psi}{dq}\Biggr)\biggl|_{a^-}=\Biggl(\frac{1}{\sqrt{M(q)}}\frac{d\psi}{dq}\Biggr)\biggl|_{a^+},
\end{equation}
which simply expresses the fact that the gradient of the wave function should be continuous at the discontinuity point. 
The continuity of the wave function itself follows from
\begin{equation}
0=\lim\limits_{\epsilon\to 0}\int_{a-\epsilon}^{a+\epsilon}\frac{1}{\sqrt{M(q)}}\frac{d\psi}{dq}\sqrt{M(q)}dq=\psi(q)|_{a^+}-\psi(q)|_{a^-}.
\end{equation}
Taking into account the above boundary conditions, together with the fact that the probability  current corresponding to $\psi^\pm$ is independent of the position $q$, we can show that  the reflection coefficient has exactly the same form as in the constant mass case, namely:
  \begin{equation}
  R=\Biggl[\frac{1-\sqrt{1-\tfrac{U_0}{E}}}{1+\sqrt{1-\tfrac{U_0}{E}}}\Biggr]^2,
  \end{equation} 
	independent of $\eta$.
	
  The Schr\"odinger equation we have used thus far was derived from the geometric interpretation of the PDEM motion. Nevertheless,  it is worthwhile to point out that in dealing with the PDEM problems, one often writes the kinetic energy part of the Hamiltonian in the form
  \begin{equation}
  \hat T=\frac{1}{4m_0}[M(q)^\nu\hat p M(q)^\mu \hat p M(q)^\kappa+M(q)^\kappa\hat p M(q)^\mu \hat p M(q)^\nu].\label{roos}
  \end{equation}
  Here $\mu$, $\nu$, and $\kappa$ are von Roos' ambiguity parameters,  which satisfy the condition $\mu+\nu+\kappa=-1$~\cite{ord}. Let us consider, for instance, the following Hamiltonian ($ V=0$)
  \begin{equation}
 \hat H=\frac{1}{2m_0} \hat p\frac{1}{M(q)}\hat p,\label{roos2}
  \end{equation}
   obtained from~(\ref{roos}) by setting $\nu=\kappa=0$.
  It can easily be verified that the Schr\"odinger equation with this Hamiltonian yields
  \begin{equation}
  \frac{d^2\psi}{dq^2}+\frac{2\eta}{m_0}\frac{d\psi}{dq}+\frac{2m_0E}{\hbar^2}{\rm e}^{-(2\eta/m_0)q}\psi=0,\label{quasifree4}
  \end{equation}
  which differs from equation~(\ref{quasifree1}) by the factor $2$. Now by making the change of variable $\xi=\sqrt{2m_0^3E/\hbar^2\eta^2}{\rm e}^{-\eta q/m_0}$, we end up with
  \begin{equation}
  \xi^2\psi^{''}-\xi\psi^{'}+\xi^2\psi=0.\label{bess}
  \end{equation}
 To eliminate the minus sign in front of the first derivative we further assume that $\psi=\xi\phi$. We then find that the function $\phi$ satisfies
 \begin{equation}
 \xi^2\phi^{''}+\xi\phi^{'}+(\xi^2-1)\phi=0.
 \end{equation}
 This is the well-known differential equation for the Bessel functions of order 1~\cite{abr}; thus the solution of~(\ref{bess}) is:
 \begin{equation}
 \phi(\xi)=C_1(\eta)J_1(\xi)+C_2(\eta) Y_1(\xi),
 \end{equation}
 where $J_1$ and $Y_1$ denote, respectively, the Bessel functions of the first and second kinds. It follows that the general solution of equation~(\ref{quasifree4}) is given by
 \begin{equation}
 \psi(q)=\sqrt{\frac{2m_0^3E}{\hbar^2\eta^2}}{\rm e}^{-\eta q/m_0}\Biggl[C_1 J_1\Bigl(\sqrt{{2m_0^3E}/{\hbar^2\eta^2}}{\rm e}^{-\eta q/m_0}\Bigr)+C_2 Y_1 \Bigl(\sqrt{{2m_0^3E}/{\hbar^2\eta^2}}{\rm e}^{-\eta q/m_0}\Bigr)\Biggr].\label{fun3}
 \end{equation}
 Besides the fact that this wave function describes a completely different quantum behavior as compared to the one obtained above in~(\ref{fun1}), here the smooth transition to the constant mass case cannot be made. Notice that this transition was possible in~(\ref{fun1}) because of  the multiplicative property of the exponential function, which does not hold for the Bessel functions. In particular, the explicit form of the probability current density corresponding to~(\ref{fun3}) turns out to be  quite involved.
 \section{Fermi gas of damped-antidamped particles and effective mass tensor\label{sec3}}
 In this section we generalize the concept of the Fermi gas to the case where the  particles, which we suppose identical with mass $m$ (we omit for now the subscript 0),  are individually subjected to the effect of three forces;  each of these is  proportional to the square of the component of the mechanical velocity along the corresponding coordinates axis of the three-dimensional configuration space, that is,
 \begin{equation}
  F_q=\eta v_q^2, \qquad  q\equiv x, y, z.\end{equation}

  Using the results of the previous section, we find that such a system admits the  Hamiltonian
 \begin{equation}
  H=\sum\limits_{i=1}^N\Biggl(\frac{ p_{x_i}^2}{2m}{\rm e}^{2\eta x_i/m}+\frac{p_{y_i}^2}{2m}{\rm e}^{2\eta y_i/m}+\frac{p_{z_i}^2}{2m}{\rm e}^{2\eta z_i/m}\Biggl),\label{fermi1}
 \end{equation}
where $N$ is the number of particles. Accordingly, we may  regard the gas as a system of particles in a medium where each particle  is characterized by the diagonal effective mass tensor ${\rm diag}(m{\rm e}^{-2\eta x/m},m{\rm e}^{-2\eta y/m},m{\rm e}^{-2\eta z/m})$.

 We   assume that the gas is in thermal equilibrium at temperature $k_B T=1/\beta$ ($k_B$ denotes Boltzmann's constant) and is completely contained in  the cube $-L\leq x,y,z \leq L$. This is the only acceptable form of the intervals, otherwise the symmetry with respect to the change of the sign of the force will be violated.   Under the latter assumptions the classical partition function reads
\begin{eqnarray}
\mathcal Z&=&\frac{1}{N!h^{3N}}\prod\limits_{k=1}^{3N}\int\limits_{-L}^L dq\int\limits_{-\infty}^{\infty}  dp\exp\Bigl\{-\frac{\beta p^2}{2m}\ {\rm e}^{2\eta q/m}\Bigr\}\nonumber\\
&=&\frac{1}{N!h^{3N}}\Bigl(\frac{2\pi m}{\beta}\Bigr)^{3N/2}\prod\limits_{k=1}^{N}\overbrace{\Biggl(\int\limits_{-L}^L dq \ {\rm e}^{-\eta\label{z} q/m}\Biggr)^3}\limits^{ V(\eta)}.\end{eqnarray}
Clearly, $V(\eta)$ represents the  volume in the 3D curvilinear space with the metric tensor ${\rm diag}({\rm e}^{-\eta x/m},{\rm e}^{-\eta y/m},{\rm e}^{-\eta z/m})$, which justifies once more the use of the geometric interpretation of the motion. The  integral in~(\ref{z}) can easily be carried out, yielding (for $\eta>0$)
\begin{equation}
\mathcal Z=\frac{1}{N!h^{3N}}\Bigl(\frac{2\pi m}{\beta}\Bigr)^{3N/2}\Biggl[\frac{2m}{\eta}\sinh\Bigl(\eta L/m\Bigr)\Biggr]^{3N}.
\end{equation}
Hence, by Stirling's formula, the free energy of the system can be written as
\begin{eqnarray}
F&=&-k_B T\ln\mathcal Z\nonumber\\
&=&-k_BTN\Biggl\{\frac{3}{2}\ln T+\ln \Biggl[\frac{1}{N}(2m/\eta)^3\sinh^3\Bigl(\eta L/m\Bigr)\Biggr]+\ln\frac{(2\pi mk_B)^{3/2} e}{h^3}\Biggr\}.
\end{eqnarray}
Knowing the explicit form of the free energy, one can calculate the diverse thermodynamical quantities characterizing the gas. In particular, the pressure turns out to be equal to
\begin{equation}
P=-\Bigl(\frac{\partial F}{\partial V}\Bigr)_{TN}=\frac{ N k_B T }{V(\eta)},
\end{equation}
with \footnote{If we suppose that the gas is contained in the region of space defined by $\{-L_1<x< L_1, -L_2<y< L_2,-L_3<z< L_3\}$, then the volume is $(2m /\eta)^3 \sinh(\eta L_1/m) \sinh(\eta L_2/m)\sinh(\eta L_3/m)$.}
\begin{equation}
V(\eta)=\Biggl[\frac{2m}{\eta}\sinh\Bigl(\eta L/m\Bigr)\Biggr]^3.
\end{equation}
Thus
\begin{equation}
PV(\eta)=Nk_B T,
\end{equation}
implying that at constant temperature, the pressure  decreases as $\eta$ increases (see discussion bellow). The chemical potential, on the other hand,  reads
\begin{equation}
\mu = \Bigl(\frac{\partial F}{\partial N}\Bigr)_{TV}
= 3 k_B T\ln\Biggl[\frac{h  \eta N^{1/3}}{2\sqrt{2\pi k_B m^3 T}\sinh\Bigl(\eta L/m\Bigr)}\Biggr].\end{equation}

In a similar way we find that the internal energy of the gas is
\begin{equation}
U=-T^2\Bigl(\frac{\partial F/T}{\partial T}\Bigr)_{VN}=\frac{3}{2}Nk_B T,
\end{equation}
which does not depend on $\eta$. As a direct consequence, we  conclude that the specific heat is equal to $C_V=\frac{3}{2}N k_B$. Hence,  although the pressure is inversely proportional to $\eta$, the classical analysis reveals no dependence of the internal energy and the specific heat on the applied force,  in contrast to what one might expect. As we shall see bellow the quantum mechanical treatment of the problem does resolve this issue.

To achieve this aim, we first look for the discrete energy levels of the gas, assuming that the potential is zero within the region of space containing the particles and is infinite outside it. The wave function of every particle of the gas is simply given by the  product of the wave functions associated with the one-dimensional motion in each coordinates axis, that is
 \begin{equation}
 \Psi(x,y,z)=\psi(x)\psi(y)\psi(z),\label{tpt}
 \end{equation}
  where
 \begin{equation}
 \psi(q)=C_1(\eta)\exp\Biggl\{i\sqrt{\frac{2 m^3 E}{\hbar^2\eta^2}} {\rm e}^{-(\eta/m)q}\Biggr\}+C_2(\eta)\exp\Biggl\{-i\sqrt{\frac{2 m^3 E}{\hbar^2\eta^2}} {\rm e}^{-(\eta/m)q}\Biggr\}.\label{edge}
 \end{equation}
 Since this should vanish at the edges of the interval $-L\leq q\le L$, it follows that
 \begin{equation}
 \exp\Biggl\{2i\sqrt{\frac{2 m^3 E}{\hbar^2\eta^2}}{\rm e}^{(\eta/m)L} \Biggr\}=\exp\Biggl\{2i\sqrt{\frac{2 m^3 E}{\hbar^2\eta^2}} {\rm e}^{-(\eta/m)L}\Biggr\},
 \end{equation}
   which yields  the quantized energy levels ($q\equiv x,y,z$)
   \begin{equation}
   E_{n_q}=\Biggl(\frac{\hbar^2\eta^2}{8m^3}\Biggr)\frac{\pi^2 n_q^2}{\sinh^2\Bigl(\eta L/m\Bigr)}, \qquad n_q=1,2,3,\cdots
   \end{equation}
  Hence the eigenenergies corresponding to~(\ref{tpt}) read
  \begin{equation}
  E_{n_x n_y n_z}=\frac{\hbar^2\eta^2\pi^2}{8m^3\sinh^2\Bigl(\eta L/m\Bigr)}(n_x^2+n_y^2+n_z^2). \label{eigen}
  \end{equation}
The wave function~(\ref{edge}), in turn, takes the form (we drop the subscript $q$)
\begin{equation}
\psi_n(q)=C^{(n)}(\eta)\Biggl\{\exp\Biggl(-i\sqrt{\frac{2 m^3 E_{n}}{\hbar^2\eta^2}} {\rm e}^{-(\eta/m)q}\Biggr)-\exp\Biggl(i\sqrt{\frac{2 m^3 E_{n}}{\hbar^2\eta^2}} ({\rm e}^{-\eta q/m }-2{\rm e}^{-\eta L/m })\Biggr)\Biggr\}.
\end{equation}
The constant $C^{(n)}(\eta)$ can be determined from the normalization condition
\begin{equation}
\int\limits_{-L}^L|\psi_{n}(q)|^2 {\rm e}^{-\eta q/m } dq=1.
\end{equation}
After some algebra, one obtains
\begin{eqnarray}
|C^{(n)}(\eta)|&=&\sqrt{\frac{\eta}{2m}}\Biggl\{2 \sinh\Bigl(\eta L/m\Bigr) -\frac{1}{2k(E_{n})}\sin\Bigl(2k(E_{n}){\rm e}^{-\eta L/m }\Bigr)\nonumber\\ &\times& \Bigl[\cos\Bigl(2k(E_{n}){\rm e}^{-\eta L/m }\Bigr)-\cos\Bigl(2k(E_{n}){\rm e}^{\eta L/m }\Bigr)\Bigr]+\frac{1}{2k(E_{n})}\cos\Bigl(2k(E_{n}){\rm e}^{-\eta L/m }\Bigr)\nonumber\\ &\times& \Bigl[\sin\Bigl(2k(E_{n}){\rm e}^{-\eta L/m }\Bigr)-\sin\Bigl(2k(E_{n}){\rm e}^{\eta L/m }\Bigr)\Bigr]\Biggr\}^{-1/2}\nonumber\\
&=&
\sqrt{\frac{\eta}{2m}}\Biggl\{2\sinh\Bigl(\eta L/m \Bigr)-\frac{1}{2k(E_{n})}\sin\Bigl[4k(E_{n})\sinh\Bigl(\eta L/m \Bigr)\Bigr]\Biggr\}^{-1/2},
\end{eqnarray}
with $k(E_{n})=\sqrt{2m^3E_{n}/\hbar^2\eta^2}$.

For completeness we also display the quantity 
\begin{eqnarray}
\mathcal P_n(L)&=&\int\limits_{-L}^L |\psi_{n}(q)|^2 dq\nonumber \\
&=&\frac{2m}{\eta}|C^{(n)}(\eta)|^2 \Biggl\{2L+\sin\Bigl(2k(E_{n}){\rm e}^{-\eta L/m }\Bigr)\nonumber\\ &\times& \Bigl[{\rm Si}\Bigl(2k(E_{n}){\rm e}^{-\eta L/m }\Bigr)-{\rm Si}\Bigl(2k(E_{n}){\rm e}^{\eta L/m })\Bigr)\Bigr]+\cos\Bigl(2k(E_{n}){\rm e}^{-\eta L/m }\Bigr)\nonumber\\ &\times& \Bigl[{\rm Ci}\Bigl(2k(E_{n}){\rm e}^{-\eta L/m }\Bigr)-{\rm Ci}\Bigl(2k(E_{n}){\rm e}^{\eta L/m }\Bigr)\Bigr]\Biggr\},
\end{eqnarray}
where
\begin{eqnarray}
{\rm Si}(x)&=&\int\limits_0^x\frac{\sin(\xi)}{\xi}\ d\xi,\\
{\rm Ci}(x)&=&\int\limits_0^x\frac{\cos(\xi)}{\xi}\ d\xi
\end{eqnarray}
are, respectively, the sine and the cosine integral functions~\cite{abr}. In figure~\ref{fig2} we display the quantity $\rho_n(q)=|\psi_n|^2\sqrt{M(q)}$ as a function of $q$ for some values of $\eta$, $L$ and $n$. We see that the probability of finding the particle is maximum in the region of space where the mass is large (see the discussion bellow). The same observation holds for $|\psi_n|^2,$ as indicated in figure~\ref{fig3}.  
\begin{figure}[htba]
{\centering{
\resizebox*{0.40\textwidth}{!}{\includegraphics{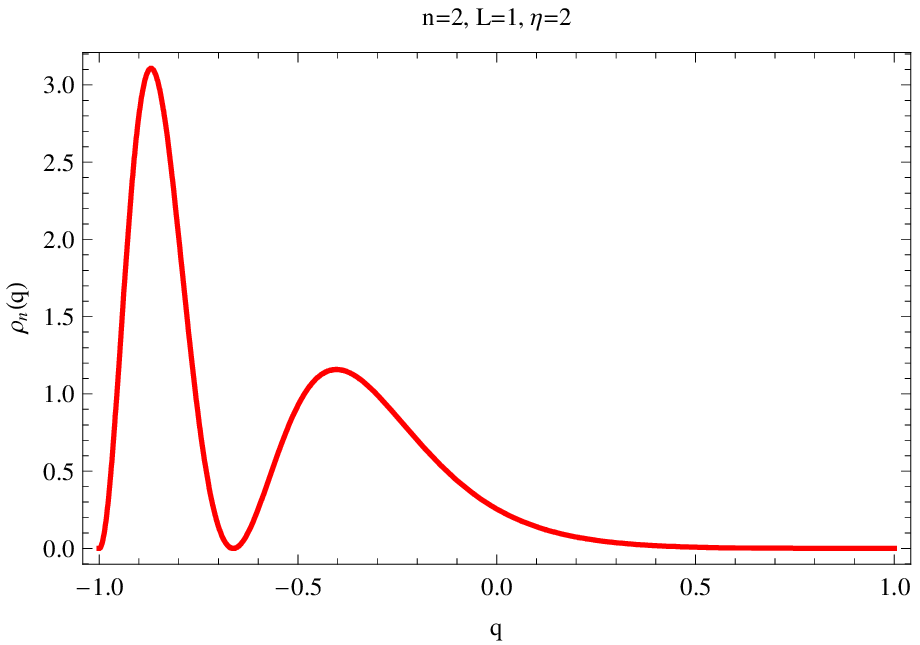}}
\resizebox*{0.40\textwidth}{!}{\includegraphics{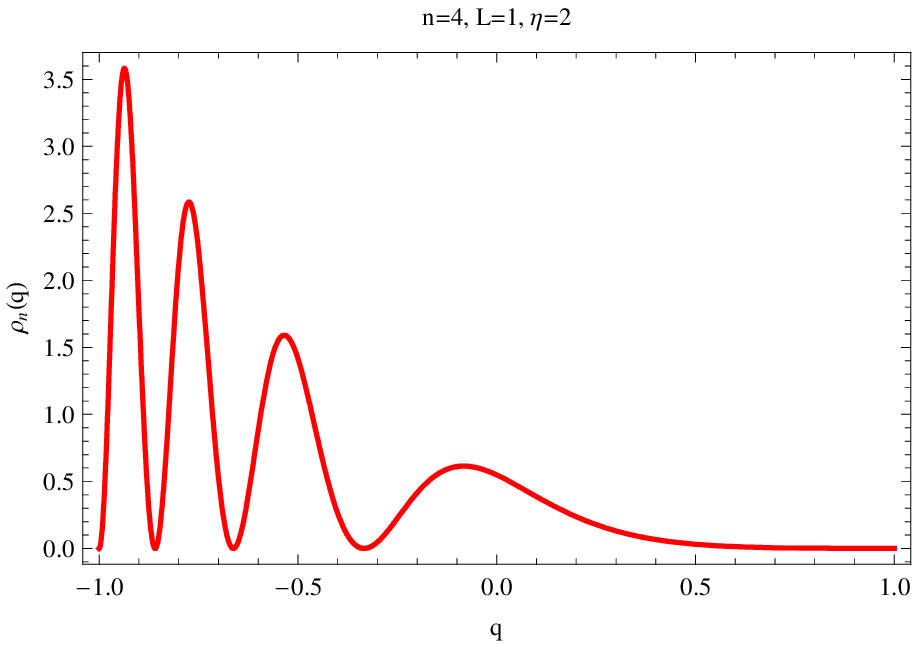}}
\resizebox*{0.40\textwidth}{!}{\includegraphics{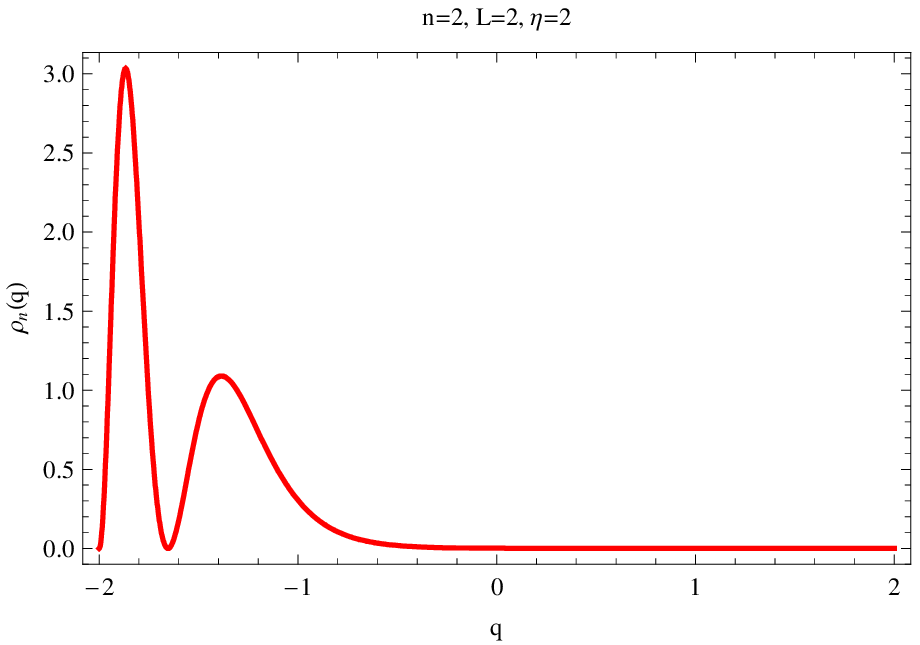}}
\resizebox*{0.40\textwidth}{!}{\includegraphics{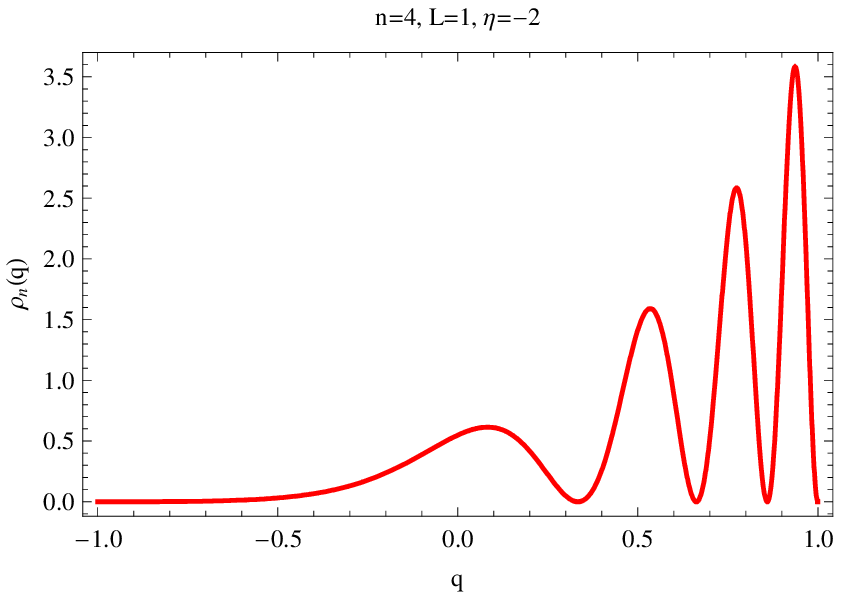}}
\par}}

\caption{\label{fig2} (Color online) The probability density $\rho_n(q)=|\psi_n|^2\sqrt{M(q)}$ as a function of position for some values of $L$, $\eta$ and $n$ with $m=\hbar=1$.  }
\end{figure}

\begin{figure}[htba]
{\centering{
\resizebox*{0.40\textwidth}{!}{\includegraphics{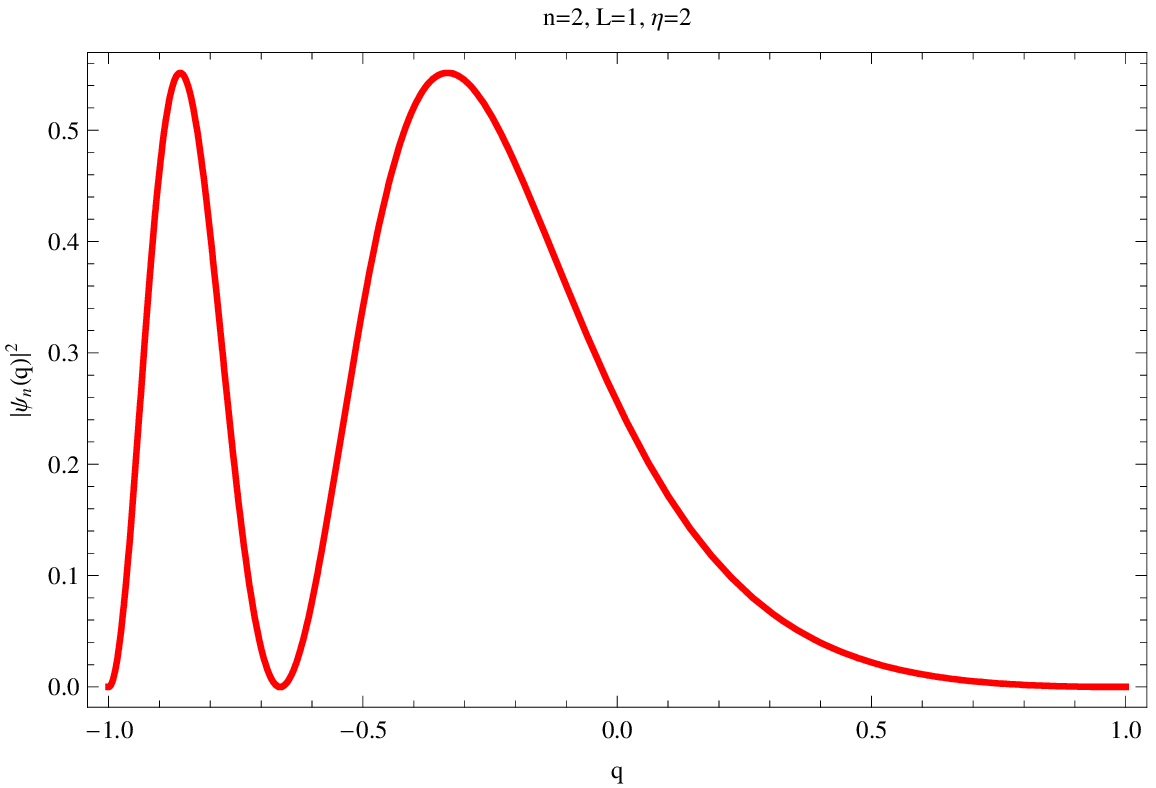}}
\resizebox*{0.40\textwidth}{!}{\includegraphics{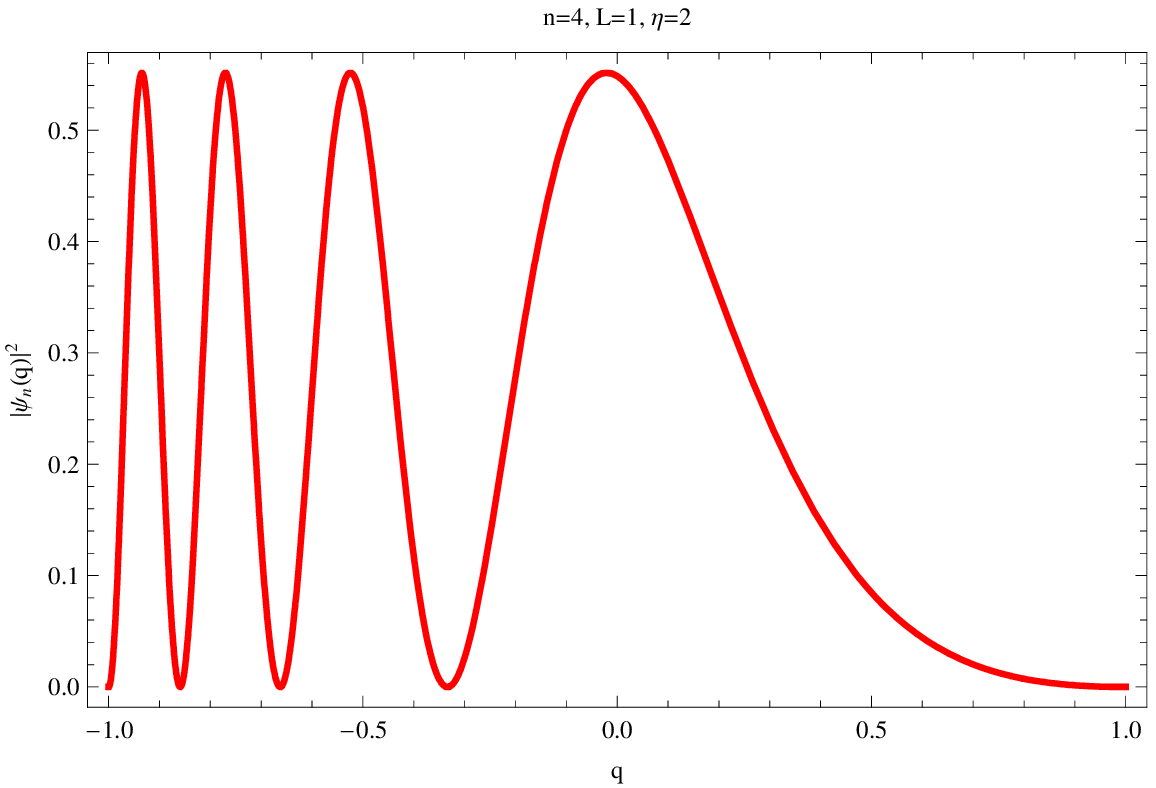}}
\resizebox*{0.40\textwidth}{!}{\includegraphics{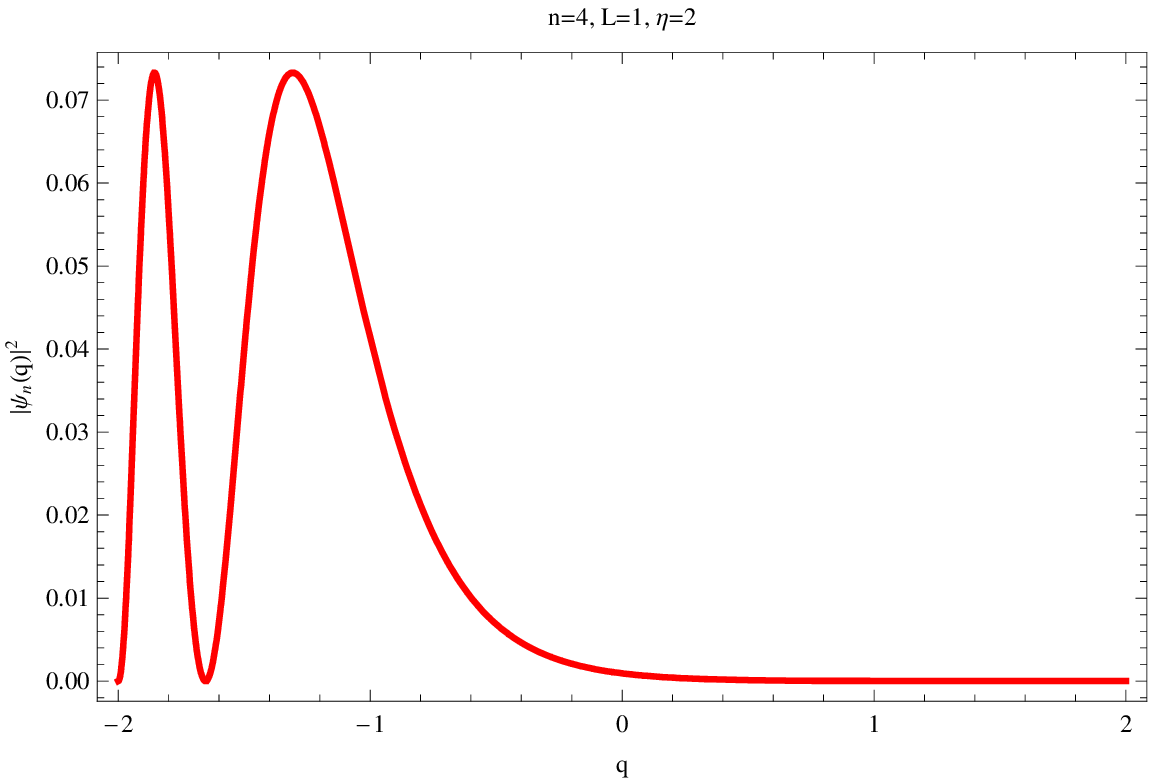}}
\resizebox*{0.40\textwidth}{!}{\includegraphics{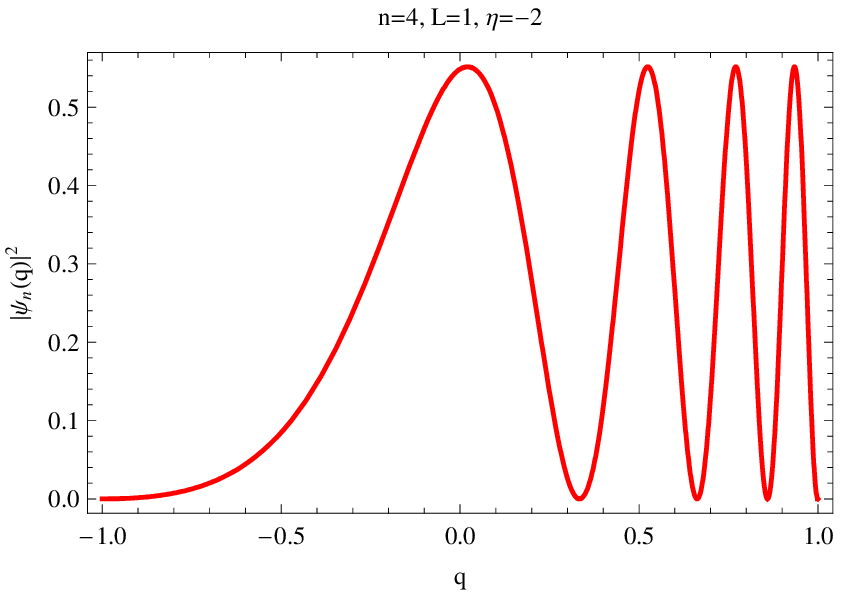}}
\par}}

\caption{\label{fig3} (Color online) $|\psi_n(q)|^2$ as a function of position for some values of $L$, $\eta$ and $n$ with $m=\hbar=1$.  }
\end{figure}

At zero temperature all the levels  are occupied up to the Fermi level, $\epsilon_F$; hence if we assume that the former are quasi-continuous~\cite{kittel}, then the density of states is equal to
\begin{equation}
dN=\frac{{8m^{9/2}} g}{\sqrt{2}\pi^2\hbar^3\eta^3}\sinh^3\Bigl(\eta L/m\Bigr)\sqrt{E}dE,\label{density}
\end{equation}
where $g$ denotes the degeneracy. Thus the total number of  particles reads
\begin{eqnarray}
N=\int\limits_0^{\epsilon_F}\frac{8{m^{9/2}} g}{\sqrt{2}\pi^2\hbar^3\eta^3}\sinh^3\Bigl(\eta L/m\Bigr)\sqrt{E}dE,
\end{eqnarray}
 from which we find that
 \begin{equation}
 \epsilon_F=\frac{\hbar^2\eta^2}{8m^3}\sqrt[3]{\Biggl(\frac{6\pi^2 N}{g}\Biggr)^2}\Biggl/\sinh^2\Bigl(\eta L/m\Bigr).\label{form}
 \end{equation}
 This equality reveals that   the Fermi energy gets smaller as $\eta$ increases, in accordance with relation~(\ref{eigen}). Now the total energy of the gas can be calculated using

 \begin{equation}
 E_{\rm tot}=\int\limits_0^{\epsilon_F} E dN,
 \end{equation}
 which yields the familiar expression
 \begin{equation}
 E_{\rm tot}=\frac{3 }{5}N \epsilon_F.
 \end{equation}
 The fact that the total energy is still proportional to the Fermi energy  explains the reason for which the pressure of the gas decreases due to the increase of $\eta$, as was found earlier via the classical partition function.

  Actually, using the results of the previous section, together with pure classical arguments, we  can explain this behavior as follows: For every particle the conservation law ($q\equiv x, y, z$)
 \begin{equation}
 \frac{d}{dt}(v_q {\rm e}^{-\eta q/m})=0,
 \end{equation}
  holds at any moment in time. Hence, for $\eta>0$ and  $q$ close to $L$, the velocity of the particle is quite large which means that it quickly leaves this region just after hitting the wall at $q=L$. On the contrary, the velocity is quite small in the neighborhood of $-L$: the particle is thus slowed down in this region, taking larger times to reach the wall at $q=-L$. Although the carried momentum is large near $q=L$, after sufficiently long time, there will be much more particles close to $-L$, as clearly indicated in figure~\ref{fig2}; hence, as a whole, the pressure of the gas drops down.

  At $T\neq0$, the number of particles is given by
  \begin{equation}
  N=\frac{{8m^{9/2}} g}{\sqrt{2}\pi^2\hbar^3\eta^3}\sinh^3\Bigl(\eta L/m\Bigr)\int\limits_0^\infty \frac{\sqrt{E}dE}{1+\exp[(E-\mu)/k_B T]}\label{nonzero},
\end{equation}
  where $\mu$ denotes the chemical potential. By introducing the notation $z =\exp(\mu/k_B T)$, and making the change of variable $\xi=\exp(-E/k_B T)$, we can write
  \begin{equation}
  N=\frac{{8m^{9/2}} g}{\sqrt{2}\pi^2\hbar^3\eta^3}\sinh^3\Bigl(\eta L/m\Bigr)\Biggr[z(k_B T)^{3/2} i\int\limits_0^1 \frac{\sqrt{\ln(\xi)}d\xi}{1+z\xi}\Biggr]\label{nonzero2}.
\end{equation}
Integrating by parts yields
\begin{equation}
N=-\frac{4m^{9/2} g (k_B T)^{3/2}}{\sqrt{2}\pi^{3/2}\hbar^3\eta^3}\sinh^3\Bigl(\eta L/m\Bigr){\rm Li}_{\frac{3}{2}}\bigl(-{\rm e}^{\mu/k_B T}\bigr) \label{nonzero3}.
\end{equation}
Here, ${\rm Li}_n(x)$ is the polylogarithm function, defined by
\begin{equation}
 {\rm Li}_n(x)=\frac{(-1)^{n-1}}{\Gamma(n-2)}\int\limits_0^1 \ln^{n-2}(y)\ln(1-x y)\frac{ dy}{y}.
 \end{equation}
 Then using equation~(\ref{form}) we find that the chemical potential is given in terms of the Fermi energy by
 \begin{equation}
 \mu=k_B T \ln\Biggl\{-{\rm Li}^{-1}_{\frac{3}{2}}\Bigg[-\frac{4}{3\sqrt{\pi}}\Biggl(\frac{\epsilon_F}{k_BT}\Biggr)^{3/2}\Biggr]\Biggl\},
 \end{equation}
 where the superscript designates the inverse function. Similarly, the total energy of the gas at nonzero temperature is found to be equal to
 \begin{eqnarray}
 E_{\rm tot}&=&-\frac{6m^{9/2} g (k_B T)^{5/2}}{\sqrt{2}\pi^{3/2}\hbar^3\eta^3}\sinh^3\Bigl(\eta L/m\Bigr){\rm Li}_{\frac{5}{2}}\bigl(-{\rm e}^{\mu/k_B T}\bigr).\nonumber\\
 &=&-\frac{9\sqrt{\pi}N k_B T}{2}\Biggl(\frac{k_B T}{\epsilon_F}\Biggr)^{3/2}{\rm Li}_{\frac{5}{2}}\bigl(-{\rm e}^{\mu/k_B T}\bigr).
 \end{eqnarray}
 Using the property~\cite{poly}
 \begin{equation}
 \frac{d}{dx}{\rm Li}_n(x)=\frac{{\rm Li}_{n-1}(x)}{x}
 \end{equation}
 we find that the specific heat  of the gas is
 \begin{eqnarray}
 C_V&=&\Bigl(\frac{\partial E_{\rm tot}}{\partial T}\Bigl)_V\nonumber\\
 &=&-\frac{9\sqrt{\pi}N }{2}\Biggl(\frac{k_B T}{\epsilon_F}\Biggr)^{3/2}\Biggl[\frac{5k_B}{2}{\rm Li}_{\frac{5}{2}}\bigl(-{\rm e}^{\mu/k_B T}\bigr)+\frac{\mu}{T}{\rm Li}_{\frac{3}{2}}\bigl(-{\rm e}^{\mu/k_B T}\bigr)\Biggr].
  \end{eqnarray}
  For $T\sim 0$ we may use the the approximations~\cite{temp}
  \begin{eqnarray}
  \mu&\approx&\epsilon_F\Biggl[1-\frac{\pi^2}{12}\Biggl(\frac{k_B T}{\epsilon_F}\Biggl)^2\Biggr],\\
  E_{\rm tot}&\approx&\frac{3}{5}N\epsilon_F\Biggl[1+\frac{5\pi^2}{12}\Biggl(\frac{k_B T}{\epsilon_F}\Biggl)^2\Biggr].
  \end{eqnarray}
  This yields
  \begin{equation}
  C_V\approx\frac{1}{2}\pi^2N k_B \frac{T}{\epsilon_F},
  \end{equation}
  which implies  that the specific heat of the gas increases with $\eta$.
\section{Motion in the Morse potential \label{sec4}}
In this section we propose to solve the time-independent Schr\"odinger  equation
\begin{equation}
\Biggl[-\frac{\hbar^2}{2m_0}\frac{1}{\sqrt{M(q)}}\frac{d}{dq}\frac{1}{\sqrt{M(q)}}\frac{d}{dq}+ U(q;\alpha)-E\Biggr]\psi(q)=0\label{inde}
\end{equation}
  for the Morse potential~\cite{morse}
\begin{equation}
U(q;\alpha)=A({\rm e}^{-2\alpha q}-2 {\rm e}^{-\alpha q}), \qquad A,\ \alpha >0.
\end{equation}
  We shall consider two special cases, namely, $M(q)=\exp \{\pm 2\eta q/m_0 \}$, with
\begin{equation}\eta/m_0=\alpha.
\end{equation}
	This choice  leads to exact analytical solutions as we shall see bellow. Physically, the latter condition implies that the effective distance constant of the  force (\ref{fric}) is equal to the effective range ($1/\alpha$ in lenght unit) of the Morse potential.
 
 Consider first the case $M(q)=\exp\{2\eta q/m_0\}$, and let us look for the discrete spectrum corresponding to negative values of $E$. By virtue of equation~(\ref{inde}), we can write
 \begin{equation}
 \frac{d^2\psi}{dq^2}-\frac{\eta}{m_0}\frac{d\psi}{dq}+\Biggl[\frac{2m_0E}{\hbar^2}{\rm e}^{2\eta q/m_0}-\frac{2m_0A}{\hbar^2}(1-2 {\rm e}^{\eta q/m_0})\Biggr]\psi=0. \label{pos}
 \end{equation}
 Making the change of variable
 \begin{equation}
 \xi=\frac{2\sqrt{-2m_0^3E}}{\hbar\eta}{\rm e}^{\eta q/m},
 \end{equation}
 equation~(\ref{pos}) becomes
\begin{equation} \psi^{''}+\Biggl[-\frac{1}{4}+\Biggl(\frac{2m_0^3A}{\hbar\eta\sqrt{-2m_0^3E}}\Biggr)\frac{1}{\xi}-\Biggl(\frac{2m_0^3A}{\hbar^2\eta^2}\Biggr)\frac{1}{\xi^2}\Biggl]\psi=0.\label{mim}
 \end{equation}
 We require that the function $\psi$ vanishes as $q\to\pm\infty$; in terms of the variable $\xi$,   we have to study the behavior of $\psi$ at infinity and in the neighborhood of zero. One can show that 
 \begin{eqnarray}
  \psi&\sim&\xi^{(1+\sqrt{1+4s^2})/2}\quad {\text{as}}\quad \xi\to0,\\
  \psi&\sim&{\rm e}^{-\xi/2}\hspace{1.75cm} {\text{as}}\quad \xi\to\infty,
 \end{eqnarray}
 where we have introduced the quantity
 \begin{equation}
 s=\frac{\sqrt{2m_0^3A}}{\hbar\eta}.
 \end{equation}
 Hence it is natural to seek a solution in the form
 \begin{equation}
 \psi(\xi)={\rm e}^{-\xi/2}\xi^{(1+\sqrt{1+4s^2})/2}\phi(\xi),
 \end{equation}
 where $\phi$ is a new function of $\xi$. By direct substitution into~(\ref{mim}), it can be verified that
 \begin{equation}
 \xi\phi^{''}+(1+\sqrt{1+4s^2}-\xi)\phi^{'}+[\tau-(1+\sqrt{1+4s^2})/2]\phi=0,\label{hyp}
 \end{equation} 
 with 
 \begin{equation}
 \tau=\frac{2m_0^3A}{\hbar\eta\sqrt{-2m_0^3E}}.
 \end{equation}
 Equation~(\ref{hyp}) is nothing but  the differential equation associated with the confluent hypergeometric function~\cite{abr,poly}
 \begin{equation}
 \phi=F(-n, 2\kappa+1,\ \xi),
 \end{equation}
 where
 \begin{equation}
 \kappa=\frac{1}{2}\sqrt{1+4s^2},\qquad n=\tau-\kappa.
 \end{equation}
It follows that the unnormalized eigenfunctions take the form
\begin{equation}
\psi_n(\xi)= {\rm e}^{-\xi/2}\xi^{(1+\sqrt{1+4s^2})/2} F(-n, 2\kappa+1,\ \xi).
\end{equation}
 The required conditions on the behavior of the wave function at $\pm\infty$ implies that only positive integer values of $n$ are allowed, whence 
 \begin{equation}
 -E_n=\frac{2m_0^3A^2/\hbar^2\eta^2}{\Bigl[(n+\frac{1}{2})+\frac{1}{2}\sqrt{1+\frac{8m_0^3A}{\hbar^2\eta^2}}\Bigr]^2}, \qquad n=n_{\rm min},\  n_{\rm min}+1,...
 \end{equation}
 Here, $n_{\rm min}$ denotes the minimum positive integer number for which $-E<A$. Thus, although the spectrum of negative eigenvalues corresponding to the Morse potential consists of a finite number of levels for constant masses, the chosen form of the position dependence of the mass we have adopted above renders the number of energy levels infinite.

Now if we assume that $M(q)=\exp\{-2\eta q/m_0\}$, then we have that
 \begin{equation}
 \frac{d^2\psi}{dq^2}+\frac{\eta}{m_0}\frac{d\psi}{dq}+\Biggl[\frac{2m_0E}{\hbar^2}{\rm e}^{-2\eta q/m_0}-\frac{2m_0A}{\hbar^2}({\rm e}^{-4\eta q/m_0}-2 {\rm e}^{-3\eta q/m_0})\Biggr]\psi=0. \label{pos2}
\end{equation}
By the change of variable
\begin{equation}
\xi={\rm e}^{-\eta q/m_0},
\end{equation}
we get 
\begin{equation}
-\frac{\hbar^2}{2m_0^3/\eta^2}\psi^{''}+A(\xi^2-2\xi)\psi=E\psi,\label{foo}
\end{equation}
which is the differential equation for a parabolic cylinder function. Let us further introduce the notations
\begin{equation}
\omega^2=\frac{2A\eta^2}{m_0^3}, \quad M=\frac{m_0^3}{\eta^2}, \quad \mathcal E=E+A.
\end{equation}
Hence with $\vartheta=1-\xi$, equation~(\ref{foo}) takes the form
\begin{equation}
\Bigl(-\frac{\hbar^2}{2M}\frac{d^2}{d\vartheta^2}+\frac{1}{2} M\omega^2 \vartheta^2\Bigr)\psi=\mathcal E\psi.
\end{equation}
Clearly,  the above procedure enables us to map the whole dynamics into  a simple harmonic oscillator problem in the interval $]-\infty,1]$. It follows that
the solution of~(\ref{pos2}) is
\begin{equation}
\psi_n(q)=C_n\exp\Biggl\{-\sqrt{\frac{m_0^3A}{2\hbar^2\eta^2}}\Bigl(1-{\rm e}^{-\eta q/m_0}\Bigr)^2\Biggr\} H_n\Biggl[\Bigl(1-{\rm e}^{-\eta q/m_0}\Bigr)\sqrt[4]{\frac{2m_0^3A}{\hbar\eta^2}}\Biggr],
\end{equation}
with the eigenvalues
\begin{equation}
E_n=\hbar\sqrt{\frac{2A\eta^2}{m_0^3}}(n+\tfrac{1}{2})-A.
\end{equation}
In the above,  $C_n$ is a constant and $H_n(x)$ denotes Hermite's polynomial of degree $n$; the allowed values of $n$ are those which satisfy the conditions
\begin{equation}
H_n\Bigl(\sqrt[4]{2m_0^3A/\hbar\eta^2}\Bigr)=0, \qquad -E_n<A.
\end{equation}
The former equation can be solved either geometrically or numerically. Nevertheless, using the property 
\begin{equation}
H_{2m+1}(0)=0, \qquad m=0, 1,2, \cdots,
\end{equation}
we conclude that for sufficiently large values of $\eta$, only odd values  of $n$ should be taken into account, provided $-E_n<A$.
\section{Summary}
Summing up, we have shown that the motion of position-dependent effective mass can be regarded as a damping-dantidamping dynamics of a constant mass under the effect of an effective potential, and vice versa. Care has to be taken when calculating the effective potentials since certain singularities may appear, which affects, in turn, the mass of the particle. We have quantized the equations of motions starting from a geometric interpretation of the motion; we found that it allows for a smooth transition to the constant-mass case, which does not hold for the momentum and mass ordering method. We have applied the obtained results to a Fermi gas of damped-antidamped particles.  The classical and the quantum treatment of this system  reveals that  the pressure and the specific heat depend on the applied forces. Finally, by solving the Schr\"{o}dinger equation in the presence of the Morse potential, we deduced the energy spectra and the eigenfunctions of the bound states in the case of a particle with exponentially decreasing or increasing mass. We found that the number of energy levels   may become infinite for particular spatial dependence of the mass.

  \end{document}